\documentclass[twocolumn,prb,amsmath,superscriptaddress,showpacs]{revtex4}
\usepackage{graphics}
\usepackage{graphicx}
\usepackage{bm}
\usepackage{epsfig}
\usepackage{pifont}

\def\laco*{LaCoO$_3$}
\def\ehs*{$\epsilon_{\text{HS}}$}
\def\eis*{$\epsilon_{\text{IS}}$}
\def\els*{$\epsilon_{\text{LS}}$}
\newcommand{\bM}{\mathbf{M}}
\newcommand{\bp}{\boldsymbol{\phi}}
\newcommand{\bk}{\mathbf{k}}
\newcommand{\cm}{\ding{51}}
\def\kha*{Khaliullin}
\def\s*{$|s\rangle$}
\def\t*{$|t_\alpha\rangle$}
\def\se*{$s$}
\def\jo*{$j_{1/2}$}
\def\jt*{$j_{3/2}$}
\def\laco*{LaCoO$_3$}
\def\ehs*{$\epsilon_{\text{HS}}$}
\def\eis*{$\epsilon_{\text{IS}}$}
\def\els*{$\epsilon_{\text{LS}}$}
\newcommand{\bR}{\mathbf{R}}

\def\kha*{Khaliullin}
\def\s*{$|s\rangle$}
\def\t*{$|t_\alpha\rangle$}
\def\se*{$s$}
\def\te*{$t_{\alpha}$}
\def\jo*{$j_{1/2}$}
\def\jt*{$j_{3/2}$}
\newcommand{\vac}{\text{v}}
\newcommand{\bvac}{\emptyset}

\newcommand{\bi}{\mathbf{i}}

\newcommand{\benu}{\be_{\nu}}
\newcommand{\be}{\mathbf{e}}
\newcommand{\mk}{\mathbf{m}_{\mathbf{k}}}
\newcommand{\mr}{\mathbf{m}(\mathbf{r})}
\newcommand{\vp}{\varphi}
\newcommand{\br}{\mathbf{r}}

\begin{document}
\title{Spontaneous Spin Textures in Multiorbital Mott Systems}

\author{J. Kune\v{s}}
\email{kunes@fzu.cz}
\affiliation{Institute of Physics,
the Czech Academy of Sciences, Na Slovance 2,
182 21 Praha 8, Czechia}
\author{D. Geffroy}
\affiliation{Department of Condensed Matter Physics, Faculty of
  Science, Masaryk University, Kotl\'a\v{r}sk\'a 2, 611 37 Brno, Czechia}

\date{\today}
\begin{abstract}
 Spin textures in $\bk$-space arising from spin-orbit coupling in non-centrosymmetric crystals find numerous applications 
 in spintronics. We present a mechanism that leads to appearance of $\bk$-space spin texture due to spontaneous
 symmetry breaking driven by electronic correlations. Using dynamical mean-field theory we show that 
 doping a spin-triplet excitonic insulator provides a means of creating new thermodynamic phases with unique properties. 
The numerical results are interpreted using analytic calculations within a generalized double-exchange framework. 
  
\end{abstract}
\maketitle
Manipulation of spin polarization by controlling charge currents and vice versa 
has attracted considerable attention due to applications in spintronic devices.
A major role is played by spin-orbit (SO) coupling in non-centrosymmetric systems. 
As originally realized by Dresselhaus~\cite{dresselhaus55} and Rashba~\cite{rashba60},
SO coupling in a non-centrosymmetric crystal lifts the degeneracy of the Bloch states at a given
$\bk$-point and locks their momenta and spin polarizations together giving rise
to a spin texture in reciprocal space. 
This leads to a number of phenomena~\cite{manchon15} such as spin-torques in ferro-~\cite{manchon08,li15} and anti-ferromagnets~\cite{zelezny14,wadley16},
topological states of matter, or spin textures in the reciprocal space that are the basis of 
the spin galvanic effect.~\cite{wunderlich09} 
Electronic correlations alone can provide coupling between spin polarization and charge currents, e.g.,
via effective magnetic fields acting on electrons moving through a non-coplanar spin background.~\cite{nagaosa12,jonietz2010}
Wu and Zhang~\cite{wu04} proposed that SO coupling can be generated dynamically in analogy to the breaking
of relative spin-orbit symmetry in $^3$He~\cite{vollhardt}. Subsequently, an effective field theory of spin-triplet Fermi surface instabilities 
with high orbital partial wave was developed in Ref.~\onlinecite{wu07}.

Here, we present a spontaneous formation of a $\bk$-space spin texture, similar to the effect of Rashba-Dresselhaus SO coupling,
in centrosymmetric bulk systems with no intrinsic SO coupling. 
The spin texture is a manifestation of excitonic magnetism that has been proposed to take place in some strongly correlated materials.~\cite{kha13,kunes14a} 
The basic ingredient is a crystal built of atoms with quasi-degenerate singlet/triplet ground states. 
Under suitable conditions a spin-triplet exciton condensate \cite{halperin68,balents00} is formed, which may adopt a variety of thermodynamic phases 
with diverse properties~\cite{kunes15}. Several experimental realizations of excitonic magnetism have already
been discussed in the literature.~\cite{kunes14b,cao14,jain15,dey16,pajskr16}

{\it Model.} We use the dynamical mean-field theory (DMFT) to study the minimal model of an excitonic magnet -- the two-orbital Hubbard Hamiltonian at half-filling
\begin{equation}
\label{eq:model}
\begin{split}
&H=\sum_{\langle ij \rangle} H_t^{(ij)} + 
\frac{\Delta}{2}\sum_{i\sigma} \left(n^a_{i\sigma}-n^b_{i\sigma}\right)+
\sum_i H^{(i)}_{\text{int}}\\
 &H^{(i)}_{\text{int}}=
U\left(n^a_{i\uparrow}n^a_{i\downarrow}+n^b_{i\uparrow}n^b_{i\downarrow}\right)
+U'\sum_{\sigma\sigma'} n^a_{i\sigma}n^b_{i\sigma'}\\
&-J\sum_{\sigma} \left(n^a_{i\sigma}n^b_{i\sigma} 
+\gamma a_{i\sigma}^{\dagger}a_{i-\sigma}^{\phantom\dagger}b_{i-\sigma}^{\dagger}b_{i\sigma}^{\phantom\dagger}\right).
\end{split}
\end{equation}\begin{figure}
\includegraphics[width=0.8\columnwidth,clip]{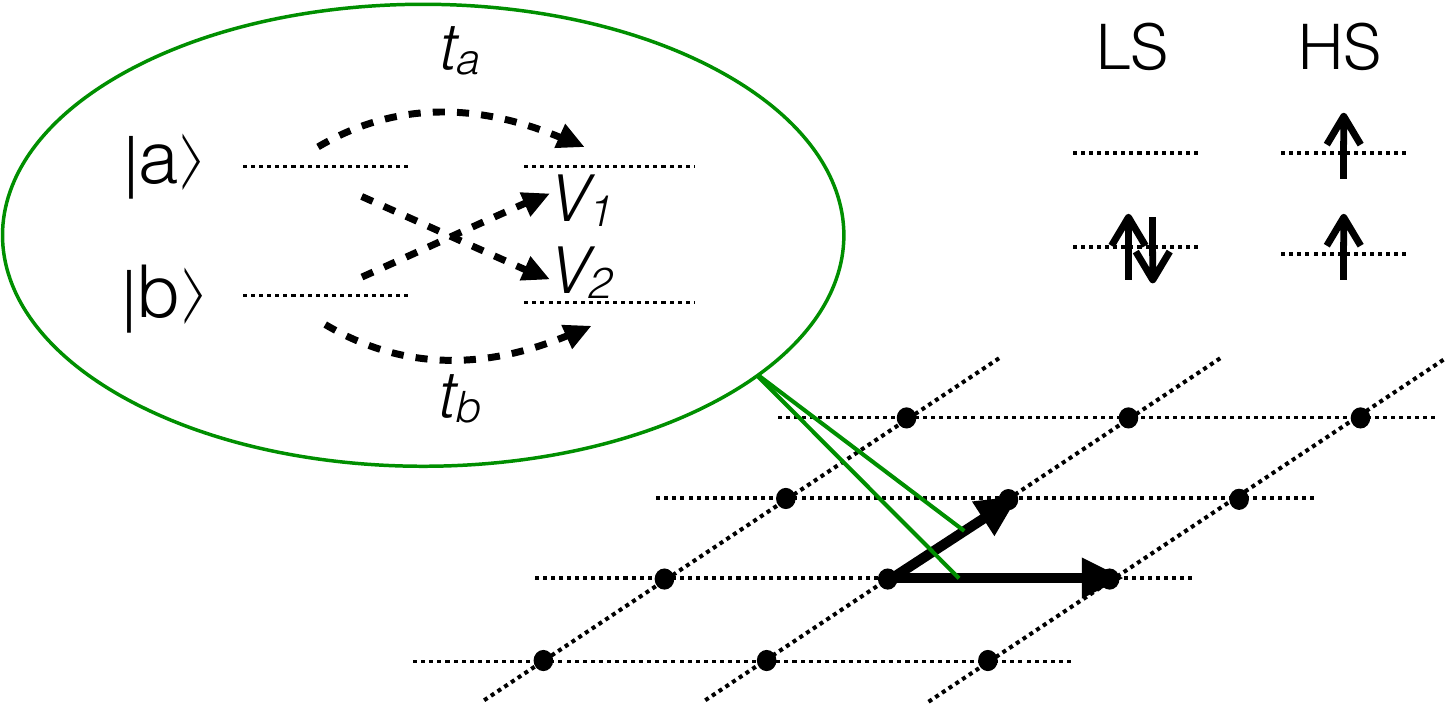} 
\caption{\label{fig:hop} The hopping processes with corresponding amplitudes on the square lattice. The parameters used in the calculations:
$t_a=0.4118$, $t_b=-0.1882$, $V_1=\pm V_2=0.05$, $\Delta=3.4$, $U=4$, $U'=2$, and $J=1$ in the units of eV.}
\end{figure}
The local part of the Hamiltonian contains the crystal-field splitting $\Delta$ between the orbitals labeled $a$ and $b$ and 
the Coulomb interaction with ferromagnetic Hund's exchange $J$. The kinetic part $H_t$ 
describes the nearest-neighbor hopping on the square lattice between the same orbital
flavors $t_{a}, t_{b}$ as well as cross-hopping between the different orbital flavors $V_1, V_2$,
see Fig.~\ref{fig:hop}.
The parameters $\Delta$ and $J$ are balanced such that the energy difference between the atomic low-spin (LS) and high-spin (HS) states
is smaller or comparable to the kinetic energy gain due to the electron delocalization.
The numerical simulations using continuous-time quantum Monte-Carlo impurity solver~\cite{werner06, alps} were performed with the density-density approximation 
for the interaction ($\gamma=0$), 
which effectively introduces a magnetic easy axis in the present model.
Analytic mean-field calculations as well as preliminary DMFT computations performed with SU(2) symmetric model~\cite{kunes15}
show only quantitative differences (e.g. reduction of the transition temperature). 
The spectral functions were obtained using the maximum entropy method.~\cite{maxent}
Technical details can be found in the Supplemental Material (SM).

Studies~\cite{kunes14a,kunes14b,kunes14c,kaneko14,kaneko15,hoshino16} performed without cross-hopping $V_{1,2}=0$ revealed formation of the exciton condensate 
below a critical temperature, which decreases with doping away from integer filling. In the strong-coupling limit the ground state wave function of 
a uniform condensate can be approximated by a product of local functions $\Pi_i |C_i\rangle$ with each
${|C\rangle=
sb_{\uparrow}^{\dagger}b_{\downarrow}^{\dagger}+\xi_1a_{\uparrow}^{\dagger}b_{\uparrow}^{\dagger}
+\tfrac{\xi_0}{\sqrt{2}}(a_{\uparrow}^{\dagger}b^{\dagger}_{\downarrow}+a_{\downarrow}^{\dagger}b^{\dagger}_{\uparrow})
+\xi_{-1}a_{\downarrow}^{\dagger}b_{\downarrow}^{\dagger}
|\text{v}\rangle}$~\endnote{The site index was dropped for the sake of simplicity.}, 
describing a local hybrid between LS and HS states with amplitudes $s$, $\xi_1$, $\xi_0$, and $\xi_{-1}$,
which provides a useful analytic reference for interpretation of the numerical results.
In the DMFT calculations we characterise the thermodynamic phases by the order parameter
${\boldsymbol{\phi}^{(i)}=\sum_{\alpha\beta}\boldsymbol{\sigma}_{\alpha\beta}\langle a_{i\alpha}^{\dagger}b_{i\beta}^{\phantom\dagger}\rangle}$,
with Pauli matrices $\boldsymbol{\sigma}$.
In addition, we evaluate the spin moment per atom $\bM$ as well as the spin density in the direct space 
$\mr$~\endnote{Without specifying the orbital shapes we only distinguish the cases with
$\mr=0$ and $\mr\neq0$.}
 and in the reciprocal space ${\mk=\sum_{\alpha\beta}\boldsymbol{\sigma}_{\alpha\beta}\langle a_{\bk\alpha}^{\dagger}a_{\bk\beta}^{\phantom\dagger}
 +b_{\bk\alpha}^{\dagger}b_{\bk\beta}^{\phantom\dagger}\rangle}$.

\begin{figure}
\includegraphics[width=\columnwidth,clip]{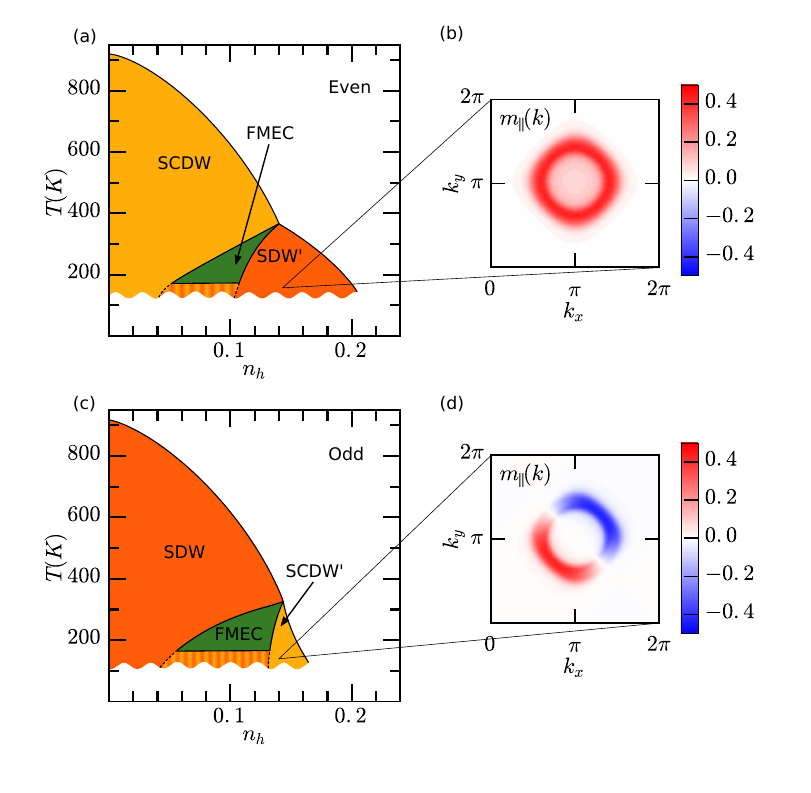} 
\caption{\label{fig:phase} (a) and (c): Phase diagrams in the doping-temperature plane for
even and odd cross-hopping, respectively.
Full lines mark continuous transitions, dotted lines mark the boundaries of phase coexistence
regions. 
(b) and (d): The spin textures at the indicated points of the phase diagrams in the units of $\mu_B (\frac{a_0}{2\pi})^2$
obtained for $n_h$=0.14 at $T$=193~K. }
\end{figure}

In Fig.~\ref{fig:phase} we show the phase diagrams of (\ref{eq:model}) as functions of temperature $T$ and hole doping $n_h$
away from $n=2$. We choose the hopping parameters so that $t_at_b<0$ which leads to a uniform $\bp$-order. 
Note that on a bipartite lattice the $t_at_b>0$ case with a staggered $\bp$-order can be mapped on the $t_at_b<0$ by the 
gauge transformation $a_i\rightarrow (-1)^ia_i$.~\cite{kunes15}
We consider two cross-hopping patterns at this point: $V_1=V_2$ (even) and $V_1=-V_2$ (odd).
The two corresponding phase diagrams share the general features inherited from the 'parent' system with no cross-hopping studied in~\onlinecite{kunes14c}.
These include the polar state with no ordered moments at low doping levels and a doping-induced transition to a different excitonic phase.
The thermodynamic phase can be distinguished by several criteria. The ferromagnetic condensate (FMEC) has the oder parameter 
of the form $\boldsymbol{\phi}= \mathbf{x}+i\mathbf{x}'$ (with non-collinear real vectors $\mathbf{x}$ and $\mathbf{x}'$), which generates
a finite uniform polarization $M_{\perp}$ perpendicular to $\boldsymbol{\phi}$. The order parameter in polar condensates can be written
as $\boldsymbol{\phi}= e^{i\varphi}\mathbf{x}$ (real vector $\mathbf{x}$ times an arbitrary scalar phase $\varphi$).
The polar condensates can be further distinguished by their time-reversal (TR) symmetry into the 
spin-density-wave (SDW; real $\bp$; breaks TR) and spin-current-density-wave (SCDW; imaginary $\bp$; preserves TR) types, introduced by Halperin and Rice,~\cite{halperin68}.
The SDW order gives rise to a finite intra-atomic spin polarization $\mr$ --higher magnetic multipole-- while the SCDW order gives rise
to intra-atomic spin current with $\mr=0$.~\endnote{Assuming the underlying orbitals are real functions.}
The preference of the undoped system for SDW or SCDW ordering on a given bond is controlled by the sign of $t_at_bV_1V_2$ and
follows the rules given in Ref.~\onlinecite{kunes14a}. Finally, we distinguish the polar phases into the primed and unprimed ones. 
The spin(current)-polarization in the unprimed phases is purely local, reflected by  $\mk=0$. The primed phase are characterized by 
appearance of $\bk$-space spin textures, $\mk\neq 0$, which in case of SCDW' phase represents global spin currents. 
The characteristics for the different phases are summarized in Table~\ref{tab:phase}.

\begin{table}
\caption{\label{tab:phase} The characteristics of different condensate phases: $M_{\perp}$ and $M_{\parallel}$ is 
magnetic moment per atom perpendicular and parallel to the order parameter $\bp$, respectively; $\mr$ and $\mk$ are
the spin densities in direct and reciprocal space, respectively. By \cm/0 we indicate that both cases may be realized (see the text). }
\begin{tabular}{|l | c | c | c | c | c | c | }
\hline
 Condensate state & $M_{\perp}$ & $M_{\parallel}$ & $\mr$ & $\mk$ & $\operatorname{Re}\bp$ & $\operatorname{Im}\bp$ \\
\hline
FMEC & \cm &\cm/0 & \cm & \cm & \cm & \cm \\
SDW & 0 & 0 & \cm & 0 & \cm & 0\\
SCDW & 0 & 0 & 0 & 0  & 0 & \cm  \\
SDW' & 0 &\cm/0 & \cm & \cm & \cm & 0 \\
SCDW' & 0 & 0 & 0 & \cm & 0 & \cm \\
\hline
\end{tabular}
\end{table}

{\it Double-exchange mechanism.} 
Observation of the spontaneous spin textures in the primed phases is our central result.
It can be understood by invoking the generalized double-exchange mechanism,
recently used by Chaloupka and Khaliullin to study ruthenates.~\cite{chaloupka16} 
Analogous to the well-known Zener double-exchange~\cite{zener55} in manganites, the exciton condensate acts as a filter for propagation of doped carriers.  
The stable phase is determined by the competition between the kinetic energy of doped carriers and the energy difference between possible condensates. 
In the strong coupling limit, propagation of a single electron
through the condensate with order parameter $\bp^{(i)}$ is described by an effective Hamiltonian (see SM for the derivation)
\begin{equation}
\label{eq:dope}
\begin{split}
H_{\text{eff}}=&\sum_{\langle ij\rangle} \left(t_s\delta_{\alpha\beta}+
\frac{1}{2}\mathbf{B}^{(ij)}\cdot \boldsymbol{\sigma}_{\alpha\beta}\right) \tilde{b}^{\dagger}_{i\alpha}\tilde{b}^{\phantom\dagger}_{j\beta}+\text{h.c.}\\
\text{with} \\
\mathbf{B}^{(ij)}=&
 \frac{it_a}{2s^2}\left(\boldsymbol{\phi}^{(j)}\wedge{\boldsymbol{\phi}^{(i)}}^*\right)
+ 
V_1^{(ji)}\boldsymbol{\phi}^{(j)}+V_2^{(ji)}{\boldsymbol{\phi}^{(i)}}^*
\end{split}
\end{equation}
and $t_s=-t_bs^2-t_a\left(1-s^2\right)$.
Here, 
$\boldsymbol{\sigma}$ are the 
Pauli matrices
and $s^2$ is the LS fraction
in the condensate. In general, the $\mathbf{B}$-fields depend in the site indices
as indicated in the brackets -
in the studied 'odd' and 'even' models the site indices are obsolete.

The $\phi$-quadratic term in (\ref{eq:dope}) describes the standard 
double-exchange interaction of the doped particle with the 
uniform background with spin polarization $\bM_{\perp}=-i\left(\boldsymbol{\phi}^*\wedge\boldsymbol{\phi}\right)/s^2$.~\endnote{$|\boldsymbol{\phi}|/s\rightarrow1$ when going from FMEC phase to a ferromagnet formed from purely HS states.}
At low doping the anti-ferromagnetic interactions between the HS states dominate, rendering
the system a polar condensate with spin-independent hopping in (\ref{eq:dope}). For some critical doping, however,
the gain in the kinetic energy of doped carriers in FMEC outweighs the cost in the 
HS-HS exchange energy and the system adopts the FMEC state.

The $\phi$-linear term in (\ref{eq:dope}), which dominates
at least close to the normal-phase boundary, appears only with finite cross-hopping
in the condensate phase.
The strong coupling calculations~\cite{kunes14a} (see SM) show that
the $V_1$ and $V_2$ contributions in (\ref{eq:dope}) cancel out, $V_1\bp+V_2\bp^*=0$,
for $\bp$ that minimizes the bond energy. On a bipartite lattice, where all bonds can be satisfied simultaneously, the $\phi$-linear term
vanishes globally allowing the SDW and SCDW phases at finite doping. 

When kinetic energy gain of the doped particles overcomes the interactions selecting the condensate type
in the undoped system, the $\phi$-linear term in (\ref{eq:dope}) becomes finite.
It has a form of an exchange field acting on bonds or equivalently acting locally in the reciprocal space, which for the two
hopping patterns considered so far reads
\begin{figure}
\includegraphics[width=0.6\columnwidth,clip]{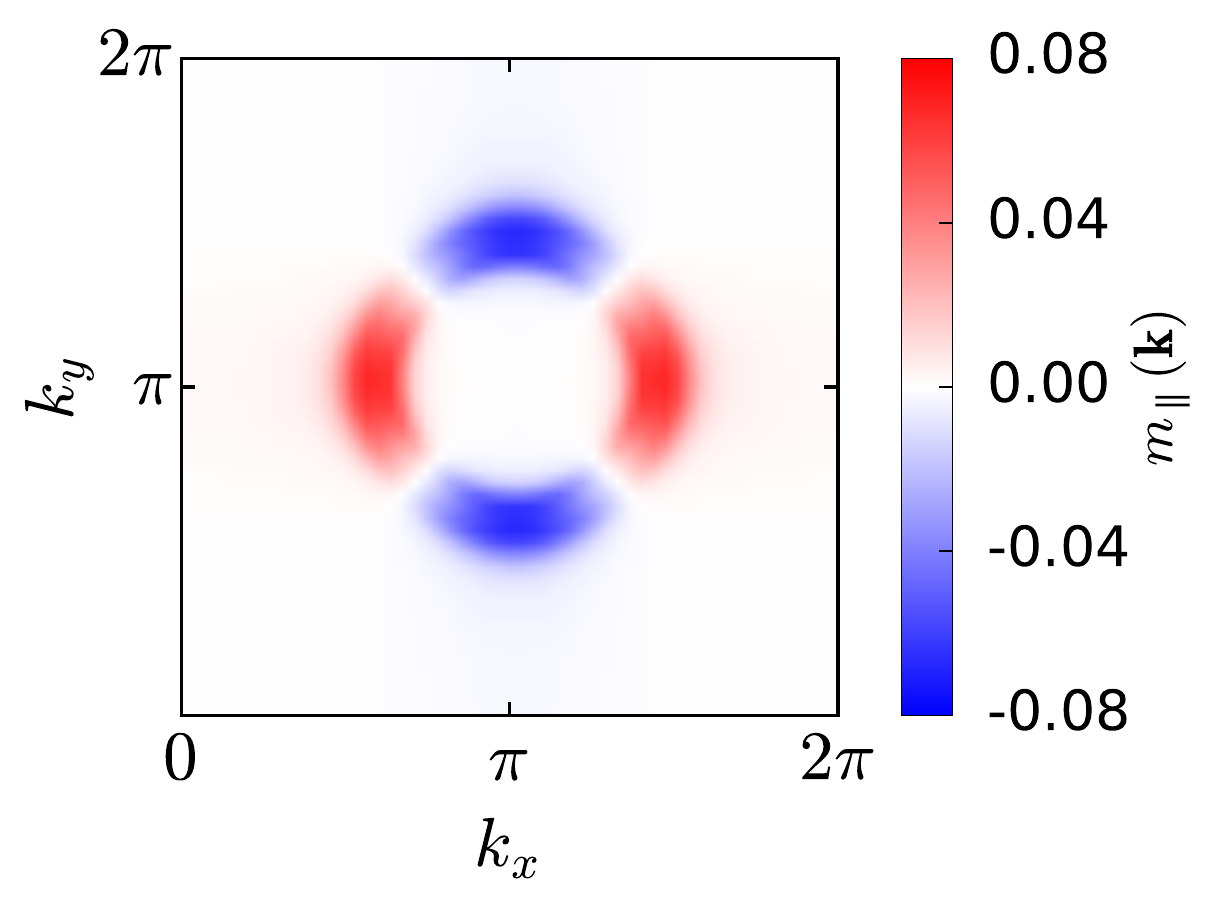} 
\caption{\label{fig:dwave} The $d$-wave spin texture in the SDW' phase of a model with even
cross-hopping of opposite signs along the $x$ and $y$ axes. The result shown here
were obtained for $n_h=0.16$ at $T$=193~K.}
\end{figure}
\begin{figure}
\includegraphics[width=\columnwidth,clip]{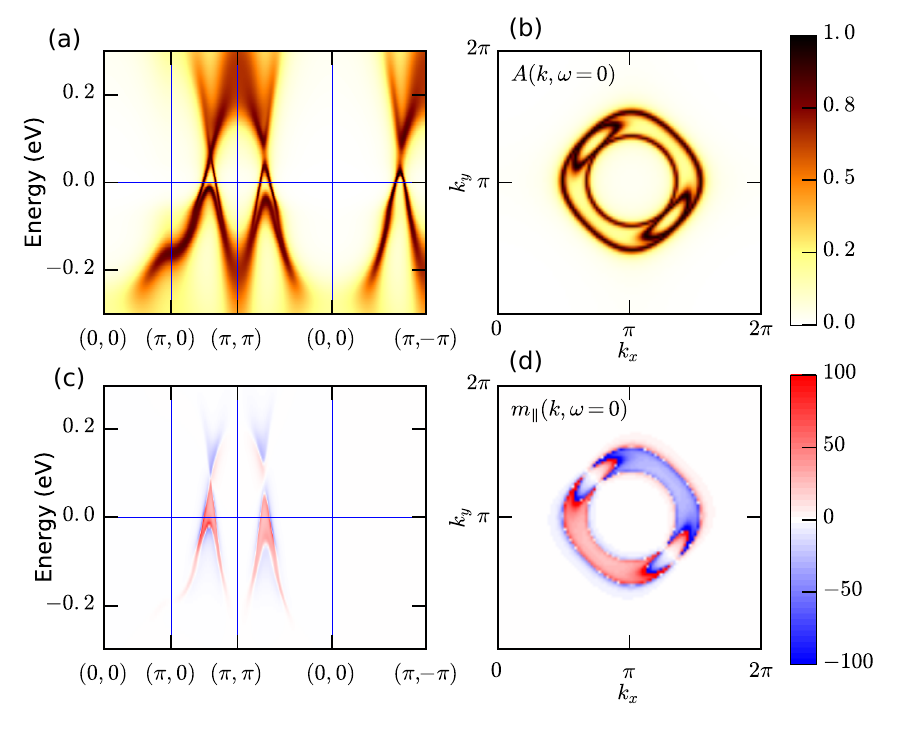}
\caption{\label{fig:density} One-particle spectral density in SCDW' phase for the same parameters as Fig.~\ref{fig:phase}d:
(a) total spectral density $A(\mathbf{k},\omega)$ along high-symmetry lines in the Brillouin zone, (b) the Fermi surface ${A(\mathbf{k},\omega=0)}$,
(c) in-plane magnetization spectral density  $m_{\parallel}(\mathbf{k},\omega)$ along the same lines as in panel (a), (d) in-plane magnetization density
at the Fermi level ${m_{\parallel}(\mathbf{k},\omega=0)}$ in the units of $\mu_B.$eV$^{-1}(\frac{a_0}{2\pi})^2$.}
\end{figure}
%
%
\begin{equation}
\label{eq:dope3}
\mathbf{B}_{\bk}=
4V_1\boldsymbol{\phi}
\begin{cases}
\cos k_x+\cos k_y & \text{SDW'}\\
 i(\sin k_x+\sin k_y)
 &\text{SCDW'}
 \end{cases} \\
\end{equation}
More generally, the $\mathbf{B}_{\bk}$ reflects the symmetry of the cross-hopping pattern.
The $s$-wave symmetry of our even cross-hopping therefore leads to an $s$-wave texture, 
Fig.~\ref{fig:phase}, with a finite $M_{\parallel}$.  Apart from strong radial localization, the $\mk$
is not qualitatively different from an approximately constant $\mk$ of normal local moment ferromagnet.
However, a $d$-wave cross-hopping, with $V$'s along the $x$ and $y$ directions having 
opposite sings, produces a $d$-wave texture, shown in Fig.~\ref{fig:dwave}, and $M_{\parallel}=0$.
We point out that without doping the $s$- and $d$-wave systems are identical, in the strong-coupling
limit, since the cross-hopping enters as a product $V_1V_2$ on each bond.~\cite{kunes14a}



The SCDW' phase is characterized by purely imaginary $\bp$ which gives rise to $\bk$-odd exchange field 
in (\ref{eq:dope3}). The odd cross-hopping pattern can be thought of as having $p_x+p_y$ symmetry, which 
is imprinted in the spin texture, shown in Fig.~\ref{fig:phase}b. There is not only no net polarization $\bM=0$, 
but the polarization is zero in every point $\mr=0$\endnote{Explicit calculation can be found in SM.} reflecting the 
TR invariance of the SCDW' state. 
In Fig.~\ref{fig:density} we analyze  spin texture in the SCDW' state in detail. The frequency-resolved contributions to $\mk$ in 
Figs.~\ref{fig:density}c,d reveal that the spin polarization comes from  a narrow energy  range around the Fermi level. Spectral functions
exhibit rather sharp quasi-particle bands around the Fermi level resembling a band structure of non-interacting system. The spin 
density, on the other hand, is quite different from that of non-interacting system. It cannot be associated
with particular quasi-particle bands but rather lives on their tails in sharply defined regions of the Brillouin zone. 

The shape of the spin texture in the SCDW' state is determined by the model parameters.  
Its collinear polarization, similar to equal combination of Rashba and Dresselhaus SO coupling~\cite{manchon15}, is picked randomly at the transition. 
The Weiss field in the SCDW and SCDW' phases, which generates local intra-atomic spin currents, can be viewed as spontaneously generated SO coupling.
The corresponding 'SO' splitting is approximately $(U-2J)|\phi|$ thus can be as large as lower units of eV. Only in the SCDW' phase the 
spontaneous SO coupling is taken to the inter-atomic scale. The equivalent of Rashba/Dresselhaus SO coupling is found in (\ref{eq:dope3}) with
the largest amplitude, in the (1,1) direction, of $4V_1|\phi|a_0$. With $|\phi|\sim 0.2-0.4$ (maximum theoretical value is $1/\sqrt{2}$),
the present cross-hopping of 50~meV, and the lattice constant $a_0$ of a few~$\AA$ the effective Rashba/Dresselhaus SO constant is 
of the order $1\times 10^{-11}$~eV m. 
\begin{figure}
\includegraphics[width=\columnwidth,clip]{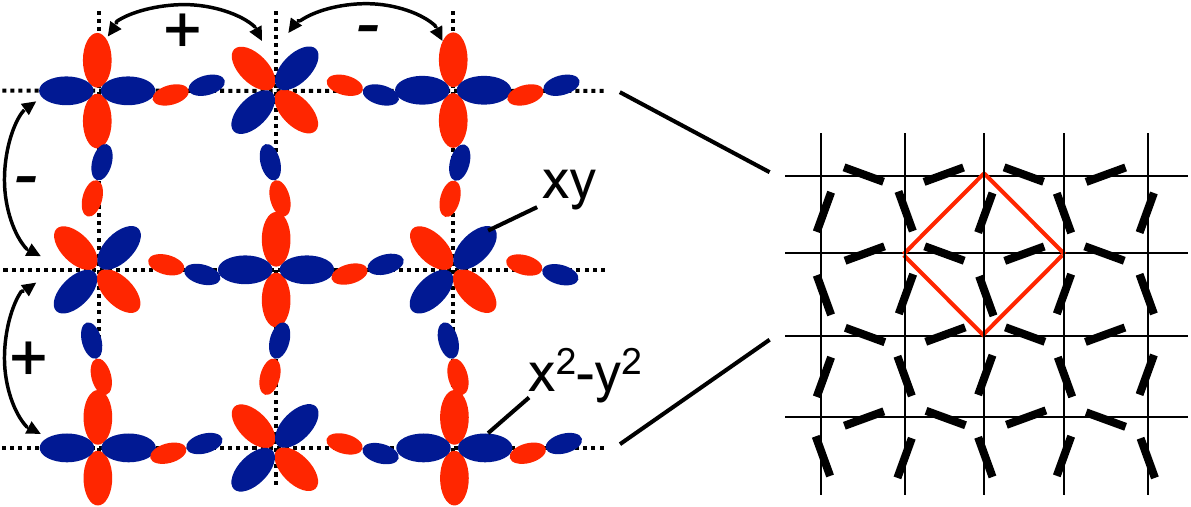} 
\caption{\label{fig:hop2} A cartoon view of the orbital pattern (left) that gives rise to
$t_a, t_b>0$ and$V_1=V_2$ on each bond with alternating signs between bonds
(only half of the orbitals is shown for sake of clarity). Zoomed out view of the 
texture on the ligand sublattice (right). Red square marks the crystallographic unit cell.  
The model can be transformed to the 'odd' cross-hopping case with a single-atom
unit cell by sublattice transformation $a_i\rightarrow(-1)^ia_i$. }
\end{figure}

{\it Realization.} To support the SDW' or SCDW' states a material:
(i) must exhibit spin-triplet polar exciton condensation, 
(ii) the local SDW or SCDW must give rise to spin-dependent hopping in Eq. \ref{eq:dope3}, and
(iii) the spin-dependent hopping must generate a global pattern spin polarization or spin currents.

Transition metal perovskites are the most discussed candidates for excitonic magnetism.~\cite{kha13,kunes14b,jain15}
The singlet-triplet quasi-degeneracy favorable for (i) is typically realized in $d^6$ configuration in octahedral geometry (Fe$^{2+}$, Co$^{3+}$, Ni$^{4+}$), $d^8$ configuration
in square planar geometry (Ni$^{2+}$), or $d^4$ configuration in octahedral geometry with strong spin-orbit coupling (Ru$^{4+}$, Os$^{4+}$, Rh$^{5+}$, Ir$^{5+}$).
Therefore we focus on models built of $d$-orbitals.

It is quite straightforward to construct the 'even' (or $d$-wave) model and thus the SDW' state from orbitals of the same parity. We focus 
on the more difficult 'odd' model and the SCDW' state. Here we have two options. First, we use the fact that only the in-plane parity is relevant. 
We can start with lattice of $3z^2-r^2$ (or $x^2-y^2$) and $z(x+y)$ orbitals. 
Breaking of the $z\leftrightarrow-z$ symmetry, e.g., by a substrate leads to the desired 'odd' cross-hopping pattern.

Second option is a model built of $x^2-y^2$ and $xy$ orbitals with more than one atom in the unit cell. In this case, the conditions (ii) and (iii) become distinct.
For example, one can obtain $V_1V_2<0$ on each bond by tilting the orbitals (oxygen octahedral in real perovskite). However, the corresponding pattern of $\mathbf{B}^{(ij)}$ has alternating signs and does not give rise to a finite $\mk$. In order, to create the desired cross-hopping pattern the inversion centered
at the atomic site has to be removed. In Fig.~\ref{fig:hop2} we show an example of such hopping pattern in Emery-like model. The diagonal hopping
amplitudes $t_a$ and $t_b$ are both negative. The cross-hopping $(V_1,V_2)$ , via tilted oxygen orbitals (induced for example by a substrate with appropriate texture),
follows the $(++),(--),(++),...$ pattern along both $x$ and $y$ directions. 
These suggestions are obviously not the only ways to realize hopping patterns favoring the SCDW' phase.

The most advanced experimental realization of the triplet-excitonic condensation is perhaps the Ca$_2$RuO$_4$~\cite{jain15} described
by the model of Khaliullin~\cite{kha13}, which is equivalent to the strong coupling limit of the present model for a special choice of parameters.
While the double-exchange mechanism is active also in ruthenates~\cite{chaloupka16}, static spin textures were not reported. 
Since the equivalents of cross- and diagonal hopping in ruthenates originate from the same $t_{2g}\rightarrow t_{2g}$ process, their ratio
is fixed and close to one. This is quite different from the present parameters with small cross-hopping.

Finally, we point out that $\bk$-space spin textures are accessible in cold atoms experiments, where the two-orbital model
may be sufficiently simple to realize. 

{\it In conclusion}, we have presented the doping of exciton condensates in systems of strongly correlated electrons as a way to generate unique states of matter. 
The generalized double-exchange mechanism in these systems can give rise to exchange fields that act on the itinerant 
electrons in the reciprocal space. The actual existence of such fields depends on the particular thermodynamic phase and crystal symmetry. In the studied model we found
a broken-symmetry state with a $\bk$-space spin texture with a symmetry of an equal combination of Rashba and Dresselhaus SO couplings. 
 
\acknowledgements
We thank G. Khaliullin, A. Hariki, L. H. Tjeng and V. Pokorn\'y for discussions, and A. Sotnikov and A. Kauch for critical reading of the manuscript. J. K. received funding from the European Research
Council (ERC) under the European Union's Horizon 2020 research and innovation programme (grant agreement No 646807) and Deutsche Forschungsgemeinschaft 
under Forschergruppe FOR1346. D. G. was supported by projects MUNI/A/1496/2014 and
MUNI/A/1388/2015 of the Masaryk University.


\newpage
\section{Supplemental Material}
\subsection{Model and computational method}
The model Hamiltonian reads
\begin{equation}
\begin{aligned}
\label{eq:2bhm}
H^{(\bi,\bi+\benu)}_{\text{t}}=&\sum_{\sigma} \left(
t_aa_{\bi+\benu\sigma}^{\dagger}a^{\phantom\dagger}_{\bi\sigma}+
t_bb_{\bi+\benu\sigma}^{\dagger}b^{\phantom\dagger}_{\bi\sigma}
 \right)+H.c.\\
+&\sum_{\sigma} \left(
V_{1}a_{\bi+\benu\sigma}^{\dagger}b^{\phantom\dagger}_{\bi\sigma}+
 V_{2}b_{\bi+\benu\sigma}^{\dagger}a^{\phantom\dagger}_{\bi\sigma}
 \right)+H.c.\\
 H^{(\bi)}_{\text{loc}}=&\frac{\Delta}{2}\sum_{\sigma} \left(n^a_{i\sigma}-n^b_{i\sigma}\right)\\
+&U
\left(n^a_{\bi\uparrow}n^a_{\bi\downarrow}+n^b_{\bi\uparrow}n^b_{\bi\downarrow}\right)+
 U'\sum_{\sigma\sigma'} n^a_{\bi\sigma}n^b_{\bi\sigma'}\\
-&J\sum_{\sigma} \left(n^a_{\bi\sigma}n^b_{\bi\sigma} 
+\gamma a_{\bi\sigma}^{\dagger}a_{\bi-\sigma}^{\phantom\dagger}b_{\bi-\sigma}^{\dagger}b_{\bi\sigma}^{\phantom\dagger}\right).\\
\end{aligned}
\end{equation}
where $\benu$ stands for the lattice vector of 2D square lattice. The DMFT calculations were performed
for the same parameters as in Ref.~\onlinecite{kunes14c}:
$U=4$, $J=1$, $U'=U-2J$, $\Delta=3.4$, $t_a=0.4118$, $t_b=-0.1882$, $V_1=\pm V_2=0.05$, $\gamma=0$ (density-density
approximation). We use eV as the energy units and give temperatures in K.

We used continuous time quantum Monte-Carlo impurity solver~\cite{werner06} modified to treat real off-diagonal hybridization functions.

The spectra were obtained with maximum entropy analytic continuation~\cite{maxent} of the self-energy. For the off-diagonal elements, 
the spectral function of which is not positive definite, we used the ansatz ${S(\omega)=S_+({\omega})-S_-({\omega})}$, where $S_+(\omega)$ and $S_-(\omega)$ 
are positive definite. We checked that  $S(\omega)$ obtained this way depends only weakly on the default model (while $S_+$ and $S_-$
are strongly default model sensitive). 

\begin{figure}[b]
\includegraphics[width=\columnwidth,clip]{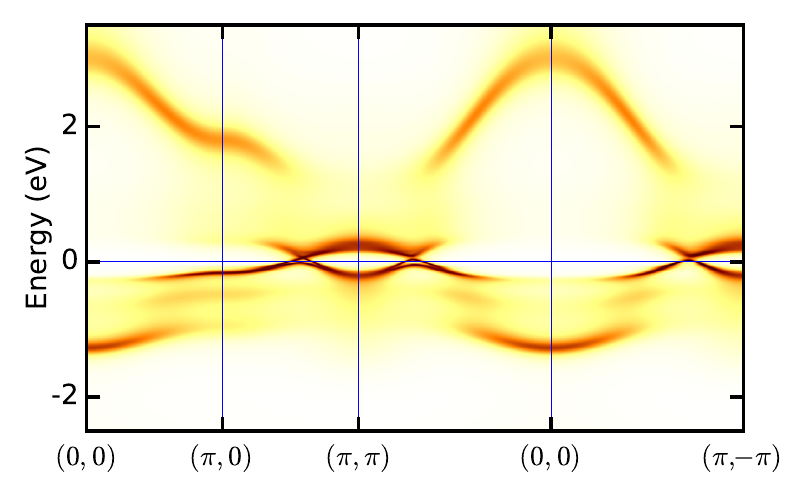}
\caption{\label{fig:spec} The spectral function over the full relevant energy range for the same parameters at in Fig.~3 of the main text.}
\end{figure}

In Fig.~\ref{fig:spec} we show the $\bk$-resolved spectral function from Fig. 4 of the article over the full energy range.
\begin{figure}
\includegraphics[width=\columnwidth,clip]{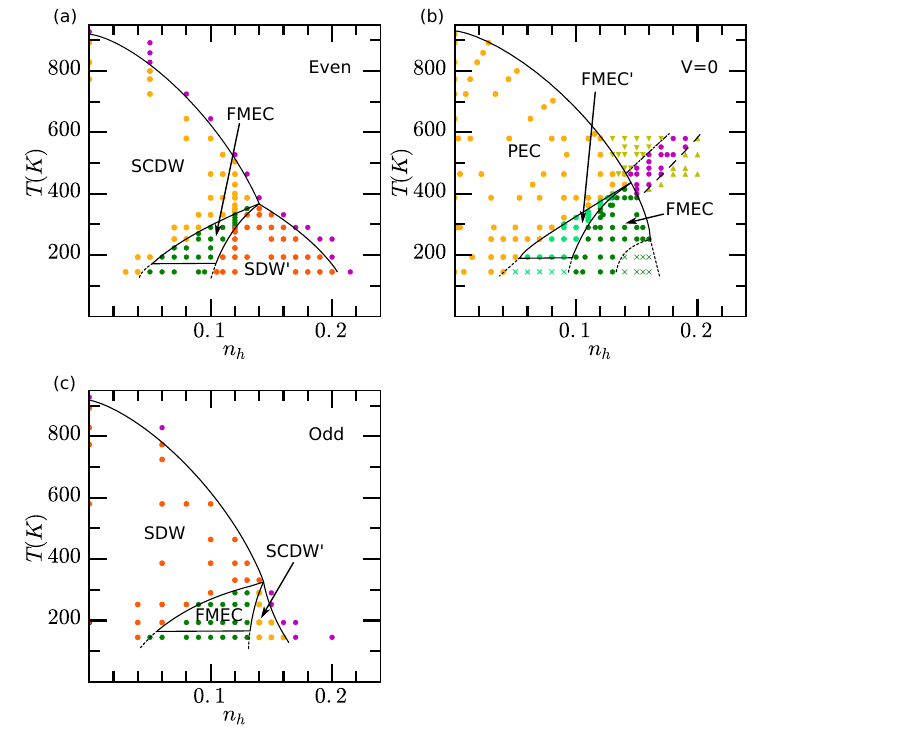}
\caption{\label{fig:dots} The input data for the Fig.~2 in the main text. The dots indicate the points at which actual calculation was performed.}
\end{figure}

\subsection{Strong coupling limit}
In the strong-coupling limit the on-site Hilbert space can be restricted to the states
\begin{equation*}
\begin{split}
&|\bvac\rangle=b_{\uparrow}^{\dagger}b^{\dagger}_{\downarrow}|\vac\rangle,\\
&|1\rangle=a_{\uparrow}^{\dagger}b^{\dagger}_{\uparrow}|\vac\rangle,
|0\rangle=\tfrac{1}{\sqrt{2}}(a_{\uparrow}^{\dagger}b^{\dagger}_{\downarrow}+a_{\downarrow}^{\dagger}b^{\dagger}_{\uparrow})|\vac\rangle,
|-1\rangle=a_{\downarrow}^{\dagger}b^{\dagger}_{\downarrow}|\vac\rangle,\\
&|\uparrow\rangle=b_{\uparrow}^{\dagger}|\vac\rangle, 
|\downarrow\rangle=b_{\downarrow}^{\dagger}|\vac\rangle,
\end{split}
\end{equation*}
where the bottom row corresponds to the doped hole states. The wave function 
of the uniform condensate can be written approximately as a product of local functions
${\Pi_i |C_i\rangle}$ with 
\begin{equation}
\label{eq:prod}
\begin{split}
&|C_i\rangle=s|\bvac_i\rangle+\xi_{-1}^{(i)}|-1_i\rangle+\xi_{0}^{(i)}|0_i\rangle+\xi_{1}^{(i)}|1_i\rangle,\\
&s^2+|\xi_1^{(i)}|^2+|\xi_0^{(i)}|^2+|\xi_{-1}^{(i)}|^2=1.
\end{split}
\end{equation}
Because the overall phase of $|C\rangle$ is physically irrelevant we will assume $s$ to be real. 
Later we will also use Cartesian representation
\begin{equation}
\label{eq:dcart}
\begin{pmatrix} \xi_{x} \\ \xi_{y}  \\ \xi_{z}
\end{pmatrix}=
\begin{pmatrix} \xi_{-1}-\xi_{1} \\ -i(\xi_{-1}+\xi_{1}) \\ \sqrt{2} \xi_{0}
\end{pmatrix}.
\end{equation}
In case of SU(2) symmetric model, the spin rotations act as SO(3) transformations on the real and imaginary parts of $\boldsymbol{\xi}$.
It is therefore always possible to make at least one of its Cartesian components zero. 
The density-density interaction, used in the numerical simulations, introduces an easy axis anisotropy which enforces the vanishing 
component to be $\xi_0$. On the level of mean-field approximation (\ref{eq:prod}), 
the solution with density-density interaction can be viewed as a solution for SU(2)-symmetric interaction for  a particular choice of $\boldsymbol{\xi}$.
This picture, however, does not extend to fluctuations around state (\ref{eq:prod}). 
  For $\xi_0=0$ the relations between the order parameter $\boldsymbol{\phi}$ and expansion coefficients in (\ref{eq:prod}) read
\begin{equation}
\begin{split} 
\phi_+^{(i)} &= \langle C_i|a_{i\uparrow}^{\dagger}b^{\phantom\dagger}_{i\downarrow}|C_i\rangle=-s{\xi_1^{(i)}}^*\\
\phi_-^{(i)}&= \langle C_i|a_{i\downarrow}^{\dagger}b^{\phantom\dagger}_{i\uparrow}|C_i\rangle=s{\xi_{-1}^{(i)}}^*\\
\phi_0^{(i)}&= \langle C_i|a_{i\uparrow}^{\dagger}b^{\phantom\dagger}_{i\uparrow}-a_{i\downarrow}^{\dagger}b^{\phantom\dagger}_{i\downarrow}|C_i\rangle=\sqrt{2}s{\xi_{0}^{(i)}}^*.
\end{split}
\end{equation}
\begin{figure}
\includegraphics[width=\columnwidth,clip]{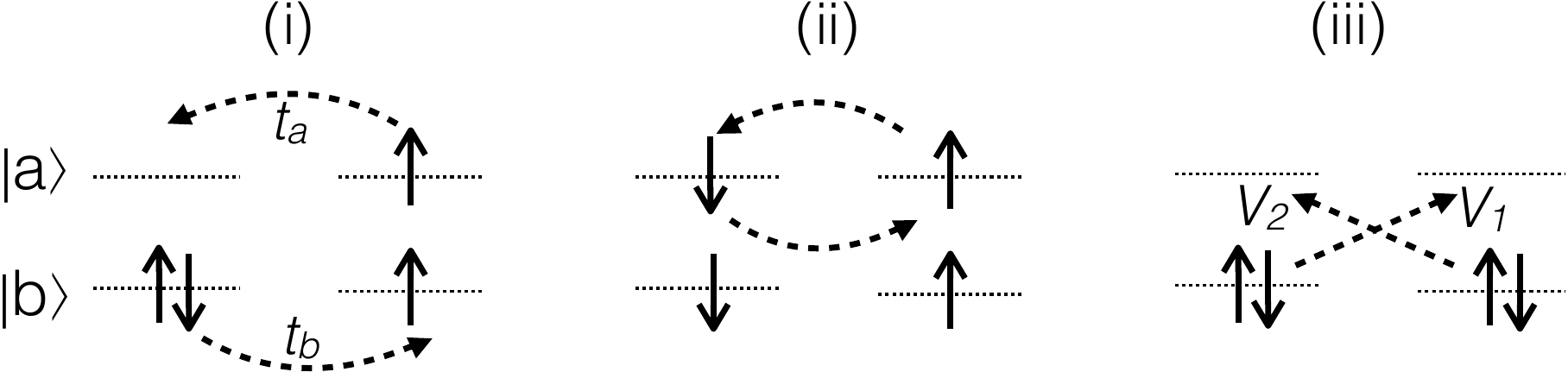}
\caption{\label{fig:s1} Nearest-neighbor hopping processes in undoped system with marked amplitudes $t_a$ and $t_b$ and
cross-hopping ${V_1=\pm V_2}$: (i) hopping of HS boson, (ii) super-exchange 
between HS states, (iii) pair creation/annihilation due to cross-hopping.}
\end{figure}

{\it Undoped case.} The ground state of the undoped system is determined by the second-order processes in hopping.~\cite{kunes15} 
In Fig.~\ref{fig:s1} we summarize the most important of these processes. The numerical results can be 
understood by looking at the signs of the different contributions to the variational energy $\langle CC|H|CC \rangle$ on the nearest-neighbor bonds 
\begin{equation*}
\begin{split}
\langle C_iC_j|H^{(i)}|C_iC_j\rangle& \sim t_at_bs^2\operatorname{Re}({\xi_1^{(j)}}^*\xi_1^{(i)}+{\xi_{-1}^{(j)}}^*\xi_{-1}^{(i)})\\
\langle C_iC_j|H^{(ii)}|C_iC_j\rangle& \sim -(t_a^2+t_b^2)\left(|\xi_1^{(i)}|^2|\xi_{-1}^{(j)}|^2+|\xi_{-1}^{(i)}|^2|\xi_{1}^{(j)}|^2\right)\\
\langle C_iC_j|H^{(iii)}|C_iC_j\rangle& \sim -V_1V_2s^2\operatorname{Re}\left(\xi_1^{(j)}\xi_{-1}^{(i)}+\xi_{-1}^{(j)}\xi_{1}^{(i)}\right).
\end{split}
\end{equation*}
The term $H^{(i)}$, which drives the transition and selects the uniform order for $t_at_b<0$, does not distinguish between the excitonic phases. The term $H^{(ii)}$, arising from nearest-neighbor anti-ferromagnetic exchange, favors the PEC phase with $|\xi_1|=|\xi_{-1}|$. 
The processes discussed so far do not distinguish the phase of the complex order parameter. 
The pair-creation term  $H^{(iii)}$ does. However, for real $V_{1,2}$ it is sensitive only to the total phase
of $\xi_1\xi_{-1}$ and depending on the sign of $V_1V_2$ it selects its value to be either 0 or $\pi$.
Both of these states can be realized with purely real $\xi_1=\xi_{-1}$ or  $\xi_1=-\xi_{-1}$. Using real $\xi_{1,-1}$ therefore
amounts, at least on the level of product state (\ref{eq:prod}), to selecting a specific direction of $\boldsymbol{\xi}$
(which will be shown to translate to the direction of spin polarization) among the possible degenerate choices.
For even cross-hopping $V_1=V_2$, $H^{(iii)}$ selects the SCDW state $\xi_1=\xi_{-1}$ ($\phi_+=-\phi_-$), while
for the odd cross-hopping $V_1=-V_2$ it selects the SDW state $\xi_1=-\xi_{-1}$ ($\phi_+=\phi_-$).

\begin{figure}
\includegraphics[width=\columnwidth,clip]{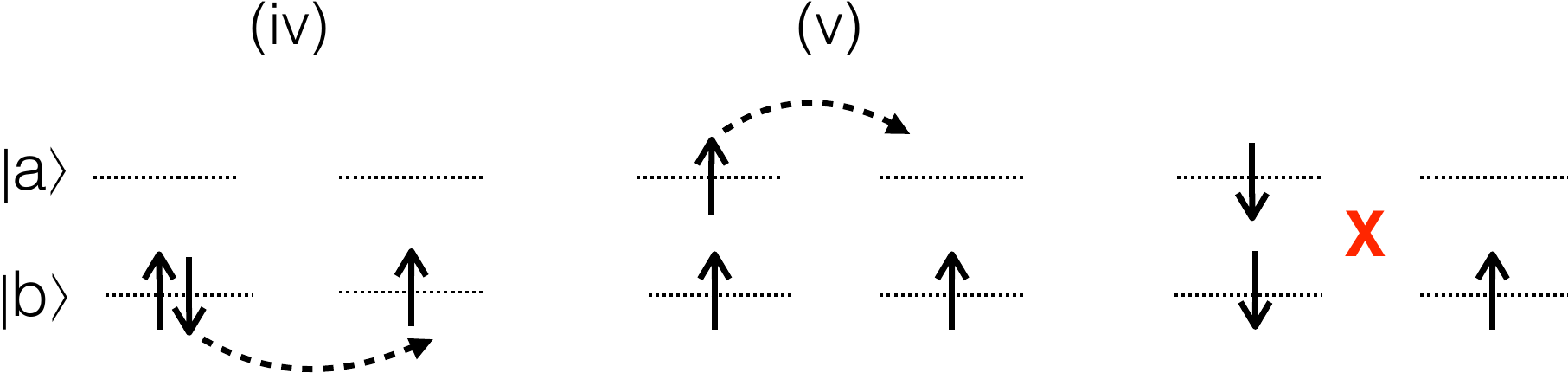}
\caption{\label{fig:s2} Nearest-neighbor processes allowing the propagation of dopes holes in system without cross-hopping:
(iv) spin-independent hole propagation, (v) spin-dependent hole propagation. }
\end{figure}

\begin{figure}
\includegraphics[width=0.7\columnwidth,clip]{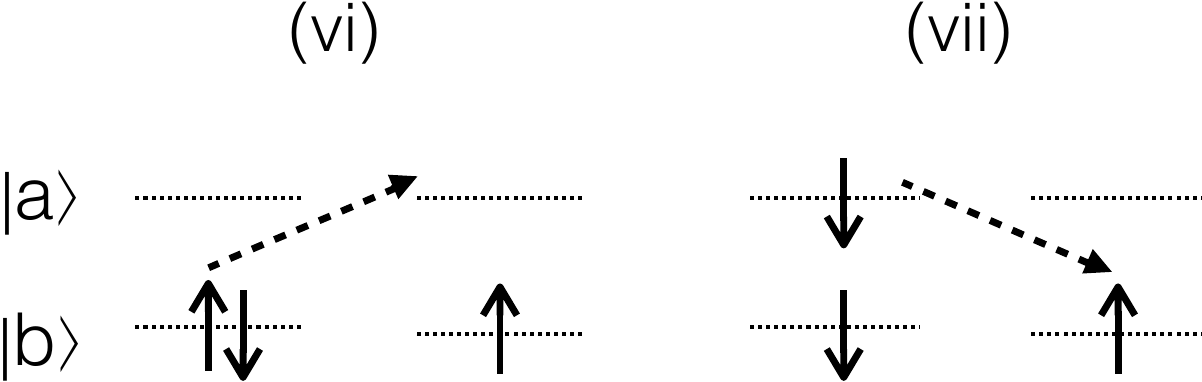}
\caption{\label{fig:s3} Additional spin-flip hopping of doped holes due to cross-hopping.}
\end{figure}

{\it Doped case.} When doped the low-energy Hilbert space contains additional states $|\uparrow\rangle$ and $|\downarrow\rangle$ that give rise to additional exchange process between
the bosonic and fermionic excitations shown in Figs.~\ref{fig:s2},\ref{fig:s3}. The simplest way to account for these processes in the low doping regime is to
compute the matrix elements describing the propagation of the doped carriers on the condensate background. This approach is well
known from the treatment of double-exchange interaction and was
recently applied in a context similar to our model.~\cite{chaloupka16}
For the sake of completeness we evaluate the matrix elements for general $\boldsymbol{\xi}$.
The contribution from hopping within the $b$-band (iv) reads
\begin{equation}
\langle \sigma_i C_j|H^{(iv)}|C_i\sigma_j\rangle= -t_bs^2.
\end{equation}
The contribution from hopping within the $a$-band (v) is spin-dependent and reads
\begin{equation}
\begin{split}
\langle \uparrow_i C_j|H^{(v)}|C_i\uparrow_j\rangle &= -t_a\left({\xi_{1}^{(j)}}^*\xi_1^{(i)}+\frac{1}{2}{\xi_{0}^{(j)}}^*\xi_0^{(i)}\right)\\
\langle \downarrow_i C_j|H^{(v)}|C_i\downarrow_j\rangle& = -t_a\left({\xi_{-1}^{(j)}}^*\xi_{-1}^{(i)}+\frac{1}{2}{\xi_{0}^{(j)}}^*\xi_0^{(i)}\right)\\
\langle \uparrow_i C_j|H^{(v)}|C_i\downarrow_j\rangle&= -\frac{t_a}{\sqrt{2}}\left({\xi_{-1}^{(j)}}^*\xi_0^{(i)}+{\xi_0^{(j)}}^*\xi_1^{(i)}\right)\\
\langle \downarrow_i C_j|H^{(v)}|C_i\uparrow_j\rangle&= -\frac{t_a}{\sqrt{2}}\left({\xi_{1}^{(j)}}^*\xi_0^{(i)}+{\xi_0^{(j)}}^*\xi_{-1}^{(i)}\right).
\end{split}
\end{equation}
The cross-hopping processes, Fig.~\ref{fig:s3}, give rise to
\begin{equation}
\begin{split}
\langle \uparrow_i C_j|H^{(vi)}|C_i\uparrow_j\rangle &= \frac{V_1^{(ji)}}{\sqrt{2}}s{\xi_{0}^{(j)}}^*\\
\langle \downarrow_i C_j|H^{(vi)}|C_i\downarrow_j\rangle& = -\langle \uparrow C|H^{(vi)}|C\uparrow\rangle\\
\langle \uparrow_i C_j|H^{(vi)}|C_i\downarrow_j\rangle&=V_1^{(ji)}s{\xi_{-1}^{(j)}}^*\\
\langle \downarrow_i C_j|H^{(vi)}|C_i\uparrow_j\rangle&=-V_1^{(ji)}Ts{\xi_{1}^{(j)}}^*
\end{split}
\end{equation}
and
\begin{equation}
\begin{split}
\langle \uparrow_i C_j|H^{(vii)}|C_i\uparrow_j\rangle &= \frac{V_2^{(ji)}}{\sqrt{2}}s\xi_{0}^{(i)}\\
\langle \downarrow_i C_j|H^{(vii)}|C_i\downarrow_j\rangle& = -\langle \uparrow C|H^{(vi)}|C\uparrow\rangle\\
\langle \uparrow_i C_j|H^{(vii)}|C_i\downarrow_j\rangle&=-V_2^{(ji)}s\xi_{1}^{(i)}\\
\langle \downarrow_i C_j|H^{(vii)}|C_i\uparrow_j\rangle&=V_2^{(ji)}s\xi_{-1}^{(i)}.
\end{split}
\end{equation}
The dynamics of the doped hole is thus described by an effective single-band Hamiltonian
\begin{equation}
\begin{split}
H_{\text{eff}}=\sum_{ij} h^{(ij)}_{\alpha\beta}\tilde{b}^{\dagger}_{i\alpha}\tilde{b}^{\phantom\dagger}_{j\beta},\\
\text{with}\ \ h_{\alpha\beta}^{(ij)}=\langle \alpha_i C_j|H|C_i\beta_j\rangle
\end{split}
\end{equation}
being the effective hopping on bond $ij$.
Using Cartesian representation (\ref{eq:dcart}) the effective hopping can be expressed in a compact form
\begin{equation}
\begin{split}
\bar{h}^{(ij)}=&-\left(t_bs^2+t_a\left(1-s^2\right)\right)\bar{I}\\
& +\frac{t_a}{4}i\left({\boldsymbol{\xi}^{(j)}}^*\wedge\boldsymbol{\xi}^{(i)}\right)\cdot\bar{\boldsymbol{\sigma}}\\
& + \frac{1}{2}\left(V_1^{(ji)}s{\boldsymbol{\xi}^{(j)}}^*+V_2^{(ji)}s\boldsymbol{\xi}^{(i)}\right)\cdot\bar{\boldsymbol{\sigma}},
\end{split}
\end{equation}
where the bar denotes $2\times2$ matrices.
For density-density interaction, which imposes the constraint $\xi_0=0$ the above equation reduces to
\begin{equation}
\bar{h}^{(ij)}=
\begin{pmatrix} 
-t_bs^2 -t_a{\xi_1^{(j)}}^*\xi_1^{(i)} & V_1^{(ji)}s{\xi_{-1}^{(j)}}^*-V_2^{(ji)}s\xi_1^{(i)} \\
 -V_1^{(ji)}s{\xi_{1}^{(j)}}^*+V_2^{(ji)}s\xi_{-1}^{(i)} & -t_bs^2 -t_a{\xi_{-1}^{(j)}}^*\xi_{-1}^{(i)}
\end{pmatrix}
\end{equation}
Hamiltonian $H_{\text{eff}}$ contains the usual spin-preserving hopping and two 'magnetic' terms proportional to $\xi$ and $\xi^2$.
The 'magnetic' terms correspond to spin-dependent hopping that can be viewed as 'magnetic' fields acting on the bonds, which give
rise to 'magnetic' fields acting locally in reciprocal space.  
With $i\boldsymbol{\xi}^*\wedge\boldsymbol{\xi}$ being the magnetic polarization of the condensate~\cite{balents00,kunes15} the $\xi^2$-term
is analogous to the Zener double-exchange interaction.~\cite{zener55} The magnetic polarization is perpendicular to the order parameter
and is not sensitive to the phase of $\boldsymbol{\xi}$. 

The $\xi$-linear term appears only for non-zero cross hopping. It gives rise to a polarization parallel to $\boldsymbol{\xi}$. 
Similar to the undoped case we can show that the mean-field ground-state energy can be minimized with real $\xi_{1,2}$. 
The kinetic energy of the doped carriers (eigenvalues of $H_{\text{eff}}(\mathbf{k})$) depends only on the amplitude of the 
off-diagonal elements of $H_{\text{eff}}$. This is, for $\xi_0=0$, proportional to
$V^2_1|\xi_{-1}|^2+V^2_2|\xi_1|^2-2V_1V_2\operatorname{Re}(\xi_1\xi_{-1})$.

\subsection{Spin-density distribution}
In the following we give an explicit formula for real space spin density $\mr$ as a function of the $\bk$-dependent one-particle density matrix.
As in the main text, we assume that the corresponding orbitals are real functions $\varphi_{\alpha}(\br)$. The summation over the
spin indices $\sigma$, $\sigma'$ is implied.
\begin{widetext}
\begin{equation}
\begin{split}
\mr&= \boldsymbol{\tau}_{\sigma\sigma'}\langle \psi^{\dagger}_{\sigma}(\br) \psi_{\sigma'}(\br) \rangle \\
&=\sum_{\bR,\bR'}\sum_{\alpha,\beta}\vp_{\alpha}(\br-\bR)\vp_{\beta}(\br-\bR')\boldsymbol{\tau}_{\sigma\sigma'}
\langle c_{\alpha\sigma}^{\dagger}(\bR) c_{\beta\sigma'}^{\phantom\dagger}(\bR')\rangle\\
&=\sum_{\bR,\bR'}\sum_{\alpha}\vp_{\alpha}(\br-\bR)\vp_{\alpha}(\br-\bR')\sum_{\bk}\cos\bk(\bR-\bR')
\boldsymbol{\tau}_{\sigma\sigma'}\langle c_{\alpha\sigma}^{\dagger}(\bk) c_{\alpha\sigma'}^{\phantom\dagger}(\bk)\rangle\\
&+\sum_{\bR,\bR'}\sum_{\alpha>\beta}\vp_{\alpha}(\br-\bR)\vp_{\beta}(\br-\bR')
\sum_{\bk}e^{-i\bk\cdot(\bR-\bR')}\boldsymbol{\tau}_{\sigma\sigma'}\langle c_{\alpha\sigma}^{\dagger}(\bk) c_{\beta\sigma'}^{\phantom\dagger}(\bk)+
c_{\beta\sigma}^{\dagger}(-\bk) c_{\alpha\sigma'}^{\phantom\dagger}(-\bk)\rangle
\end{split}
\end{equation}
\end{widetext}


\end{document}